# Quantifying magnetic field driven lattice distortions in kagome metals at the femto-scale using scanning tunneling microscopy

Christopher Candelora[1,] *, Hong Li[1,] *, Muxian Xu[1], Brenden R. Ortiz[2], Andrea Capa Salinas[3], Siyu Cheng[1], Alexander LaFleur[1], Ziqiang Wang[1], Stephen D. Wilson[3] and Ilija Zeljkovic[1,§]

[1] Department of Physics, Boston College, Chestnut Hill, MA 02467, USA

[2] Materials Science and Technology Division, Oak Ridge National Laboratory, Oak Ridge, 37831, Tennessee, USA

[3] Materials Department, University of California Santa Barbara, Santa Barbara, California 93106, USA

* Equal contribution

§Correspondence: ilija.zeljkovic@bc.edu

**Abstract**

**A wide array of unusual phenomena has recently been uncovered in kagome solids. The charge density wave (CDW) state in the kagome superconductor $AV_3Sb_5$ in particular intrigued the community – the CDW phase appears to break the time-reversal symmetry despite the absence of spin magnetism, which has been tied to exotic orbital loop currents possibly intertwined with magnetic field tunable crystal distortions. To test this connection, precise determination of the lattice response to applied magnetic field is crucial, but can be challenging at the atomic-scale. We establish a new scanning tunneling microscopy based method to study the evolution of the $AV_3Sb_5$ atomic structure as a function of magnetic field. The method substantially reduces the errors of typical STM measurements, which are at the order of 1% when measuring an in-plane lattice constant change. We find that the out-of-plane lattice constant of $AV_3Sb_5$ remains unchanged (within $10^{-6}$) by the application of both in-plane and out-of-plane magnetic fields. We also reveal that the in-plane lattice response to magnetic field is at most at the order of 0.05%. Our experiments provide further constraints on time-reversal symmetry breaking in kagome metals, and establish a new tool for higher-resolution extraction of the field-lattice coupling at the nanoscale applicable to other quantum materials.**

Introduction

Magnetic phase transitions typically lead to a small volumetric expansion anomaly. This is due to a sudden internal strain build-up caused by the spontaneous alignment of spins in order to reduce the total energy of the material. Such a spontaneous internal strain is directional and can be quantified by a magnetostriction coefficient $\lambda = dl/l$. This coefficient is usually maximized along the magnetization axis and minimized perpendicular to it, and can be controlled by applying the magnetic field to align the domains. Beyond the rotation and alignment of spins, additional crystal expansion can occur by applying an even higher magnetic field to the material with already saturated spins. The magnetostriction effect is typically small but in some cases it can reach exceptionally large values at the order of several percent, in for example $YMn_2$ [1]. To measure magnetostriction, traditional experiments rely on scattering techniques in magnetic field, which average the signal from an entire bulk sample. The presence of structural domains can however complicate the interpretation of these measurements. Measuring the local coupling between magnetic field and the lattice in nanoscale regions of the sample would be highly desirable, but this has been challenging. Here we establish a new method based on scanning tunneling microscopy to quantify small magnetic field induced lattice constant change and apply it to the intriguing example of $A$V$_3$Sb$_5$ kagome superconductors.

The family of kagome superconductors AV$_3$Sb$_5$ (A=K, Cs, Rb) attracted a tremendous attention in recent years [2–26] for unusual superconductivity, surprising density wave states and the unconventional anomalous Hall effect despite the absence of spin magnetism. Much of the effort has been focused on understanding the symmetry of the $2a_0 \times 2a_0$ charge density wave (CDW) state in the kagome plane [6–8,27–46]. Theory suggested that a chiral CDW phase composed of orbital loop flux currents could naturally explain the emergent time-reversal symmetry breaking and the anomalous Hall effect [32,47–49]. Moreover, it has been reported that externally applied magnetic field appears to control the anisotropy of the CDW signal by generating an in-plane shear strain of about 0.5% of the in-plane lattice constant [50]. Such strong connection between the crystal structure and the magnetic field could provide the elusive unifying signature associated with time reversal symmetry breaking in AV$_3$Sb$_5$. To shed light on this emerging picture, comprehensive experiments probing magnetic field driven lattice response in AV$_3$Sb$_5$ along different crystalline directions and with control of the magnetic field direction are crucial.

Here we quantify magnetostriction and piezo-magnetic response in AV$_3$Sb$_5$ (A=Cs, K) over nanoscale regions as a function of magnetic field. We introduce a new scanning tunneling microscopy (STM) based method to detect tiny changes in the lattice constants. The method substantially decreases the uncertainty of determining in-plane lattice constant variation with applied field due to the inevitable experimental artifacts that artificially distort the images, including thermal and piezo hysteresis and relaxation effects, which can be at the order of 1% in typical STM measurements. We observe that the out-of-plane lattice constant remains unaffected by external magnetic field. In all the regions studied, we find that magnetic field of a few Tesla leads to at most 0.05% change in the in-plane lattice constant change. Reversal of the magnetic field direction also leads to a similarly small effect. Our work reveals the upper bound of the magnitude of the coupling between the lattice and external magnetic field in AV$_3$Sb$_5$, thus providing further constraints on the origin of time-reversal symmetry breaking in AV$_3$Sb$_5$. It also establishes a new nanoscale method for determination of magnetic field induced structural changes using STM with better accuracy, which could be widely applied to other quantum solids.

**Results**

Kagome superconductors $A$V$_3$Sb$_5$ crystallize in the *P6/mmm* space group [2], and cleave between hexagonal Sb layers and alkali $A$ layers [6–8,42,45]. We start our experiments with the most commonly studied Cs-variant, CsV$_3$Sb$_5$. STM topographs of the Sb termination of CsV$_3$Sb$_5$ (Fig. 1b) exhibit the expected hexagonal lattice, with in-plane lattice constants $a \approx b \approx 5.4$ Å. Consistent with previous experiments [7,10], Fourier transforms (FTs) of the STM topographs show the atomic Bragg peaks **Q**$_{Bragg}$, as well as the $4a_0$ charge-stripe order and the $2a_0 \times 2a_0$ CDW in the kagome plane (Fig. 1a). Early STM experiments studying the effects of magnetic field primarily focused on the intensities of the CDW peaks as a function of magnetic field [6,8,51]. In all our measurements of AV$_3$Sb$_5$ samples, we observe no substantial change in the CDW intensity of peaks (Fig. 1c). To investigate the field-lattice coupling, here we specifically focus on the field-dependent measurements of the $\mathbf{Q}_{Bragg}$ wave vector lengths along different lattice directions.

Atomic Bragg peaks in the FTs are directly related to the lattice constant via a relation: $\mathbf{Q}_{Bragg} = \frac{4\pi}{\sqrt{3}a}$, where $a$ is the real-space lattice constant. We determine the position of each atomic Bragg peak $\mathbf{Q}_{Bragg}$ in the FT by fitting a 2D Gaussian function (Fig. 1a,d). We note that due to a small inequivalence between different piezoelectric scanner electrodes and piezo relaxation effects during scanning [52], there typically exists a small apparent difference between $|\mathbf{Q}^a_{Bragg}|$, $|\mathbf{Q}^b_{Bragg}|$ and $|\mathbf{Q}^c_{Bragg}|$ extracted from STM topographs, even for a material characterized by a perfect hexagonal lattice. However as we will show in the subsequent paragraphs, we establish a procedure such that a *relative* difference between $|\mathbf{Q}^i_{Bragg}|$ acquired in different fields can provide a robust measure of the relative lattice constant change.

For pedagogical purposes, let us examine a series of consecutively taken STM topographs over one area of the sample. As schematically depicted in Fig. 1e, topographs are acquired in either forward or backward directions (which we define as the "fast" scan direction), with tip gradually moving upwards or downwards (which we define as the "slow" scan direction). We compare the apparent $|\mathbf{Q}^i_{Bragg}|$ magnitudes in the FTs of each of these topographs (Fig. 1f-h). Our first observation is that $|\mathbf{Q}^a_{Bragg}|$, the atomic Bragg peak most closely aligned with the fast scan direction in this data set, generally shows a consistent value. On the other hand, the variation of atomic Bragg peak magnitudes along the other two directions can be substantial, at the order of 1% seen in for example $\mathbf{Q}^c_{Bragg}$ (Fig. 1f). This finding is consistent with the general expectation that drift effects will be minimized along the scan direction [52]. To examine this further, we acquire and analyze equivalent data with fast scan directions most closely aligned with $\mathbf{Q}^b_{Bragg}$ (Fig. 1g) or $\mathbf{Q}^c_{Bragg}$ directions (Fig. 1h). The data and the analysis shows that $|\mathbf{Q}^i_{Bragg}|$ measured along the fast scan direction consistently shows minimal variations, while the other two can show substantial fluctuations.

To further illustrate this issue, we turn to the comparison between STM topographs acquired at zero field and those at finite magnetic field **B** (Fig. 2). To quantify the apparent relative in-plane lattice constant change from this data, we define a coefficient: $\lambda^{B}_{in-plane} = \frac{|\mathbf{Q}^i_{Bragg}(B)| - |\mathbf{Q}^i_{Bragg}(0T)|}{|\mathbf{Q}^i_{Bragg}(0T)|}$. Using this formula, we calculate and plot magnetostriction coefficients over an identical area of the sample for several representative fields (Fig. 2a-c), with the fast scan direction approximately along $\mathbf{Q}^a_{Bragg}$ wave vector. We observe that the change in $|\mathbf{Q}^a_{Bragg}|$ after the initial scan is 0.05% or less for all fields (Fig. 2a-c). In contrast, the apparent change in $|\mathbf{Q}^b_{Bragg}|$ and $|\mathbf{Q}^c_{Bragg}|$ can be as large as 0.5% (Fig. 2a-c). This is summarized in the scatter plot of average $\lambda^{B_z}_{in-plane}$ vs $B_z$ in Fig. 2d. Importantly, the equivalent experiment performed over the same region of the sample, but with the fast scan direction along $\mathbf{Q}^b_{Bragg}$ wave vector yields striking differences (Fig. 2e). In particular, $|\mathbf{Q}^b_{Bragg}|$ that is now along the fast-scan direction shows less than 0.05% change, in striking difference to the large $\lambda^{B_z}_{in-plane}$ apparent in the rotated scan frame. This data and analysis demonstrate that the anomalously large coefficient measured in Fig. 2d is not an intrinsic feature of the sample, but instead an experimental artifact likely due to thermal or piezo drift. This cautions the use of STM data quantitative analysis and highlights the crucial need for examining a sequence of rotated scan frames for robust detection of lattice constant changes.

For these reasons, in the following field-dependent experiments, we acquire three sets of topographic scans, with fast scan directions aligned to each of the atomic Bragg peaks within a few degrees; we calculate and use $|Q^i_{Bragg}|$ values extracted from STM topographs for which the Bragg peak is most closely aligned with the fast scan direction. When magnetic field is applied along the z-direction, we find that average $\lambda^{B_z}_{in-plane}$ along $Q^a_{Bragg}$ and $Q^c_{Bragg}$ directions is at most 0.01% (Fig. 3b, Supplementary Figure 1). We note that $\lambda^{B_z}_{in-plane}$ along $Q^b_{Bragg}$ is slightly larger with more substantial error bars due to the quality of the particular data set. We then conduct the equivalent measurement on another sample, using a different microscope, to apply ± 2T in-plane magnetic field as schematically shown in Fig. 3a. We again find that majority of data points are around 0.01% and all of them below 0.05% (Fig. 3d).

To investigate if magnetic field oriented in opposite directions can induce more unusual types of deformations [50], we directly compare the lattice response when magnetic field direction is reversed (Fig. 3d-f). Referred to as a piezo-magnetic response linear in magnetic field, this effect will in principle result in different lattice constants for **B** vs –**B**, magnetic fields of the same magnitude but applied in opposite directions [53]. To quantify this type of anti-symmetric response, we define a piezo-magnetic coefficient: $\mu^B_{in-plane} = \frac{2(|Q^i_{Bragg}(B)|-|Q^i_{Bragg}(-B)|)}{|Q^i_{Bragg}(B)|+|Q^i_{Bragg}(-B)|}$ , which yields the change in each in-plane lattice constant after magnetic field direction is reversed. From the analysis of our data, we find that $\mu^{B_z}_{in-plane}$ is less than 0.05% in $CsV_3Sb_5$ for all in-plane lattice directions (Fig. 3e). We note that we performed similar experiments on cousin $RbV_3Sb_5$ (Fig. 3c, f) and $KV_3Sb_5$ samples for one scan rotation (Supplementary Figure 2), and arrive at the same conclusion of minimal in-plane lattice constant change along the fast-scan direction driven by magnetic field.

Different from a polycrystalline sample, which can be considered to be macroscopically isotropic and thus with a field-aligned magnetostriction coefficient independent of the measuring direction, magnetic field lattice response in a single crystal could also vary for different crystalline axes. STM cannot directly measure the lattice constant along the c-axis direction, but as we will show, it can be sensitive to changes in the overall sample thickness with sub-Angstrom resolution. In particular, we proceed to use the change in the STM tip height $\Delta Z$ in the tunneling regime as the proxy for the total sample thickness change (Fig. 4a). We use specific impurities in the field-of-view as markers for tracking an identical region of the sample and extract the tip height difference $\Delta Z$ as a function of magnetic field. Similar procedure was previously applied to uncover a large out-of-plane lattice response in magnetically ordered systems [54] and here we use it to investigate if subtle effects may be present in non-magnetic kagome metals. It is important to note that while the magnetic field is being changed, tip is withdrawn (negative voltage is applied to piezoelectric scanner electrodes) so that it is not in feedback, and no piezoelectric "walker" steps are taken to bring the entire scanner away from the sample surface.

Overall, we observe a nanometer scale effect (Fig. 4b,c). For example, applying magnetic field of 4 T along the z-direction leads to an apparent tip change of about 7 nm for $CsV_3Sb_5$ (Fig. 4b, c) and a change of about 8 nm for $RbV_3Sb_5$ (Fig. 4b, g). To investigate the expansion of the entire microscope setup, including the sample holder, we study the tip height variation in the tunneling regime over a wide range of different materials, including cleaved bulk single crystals and thin films, of different thicknesses (Fig. 4b). All of them show a linear evolution in magnetic field, with varying slopes. Helical antiferromagnet $YMn_6Sn_6$ shows the largest slope (Fig. 4b), which may not be surprising given its magnetic nature with the expected continuous field-induced spin tilting up to a much higher magnetic field than used here [55–59]. Most other materials fall along a similar line with a lower slope. To approximate the background expansion of the sample holder, we calculate the smallest slope and subtract it from our raw measurements. Taking into account the thickness of the $CsV_3Sb_5$ sample used in the experiment to be 0.16 ± 0.02 mm, the resulting normalized value of the out-of-plane magnetostriction coefficient $\lambda_{out-of-plane}^{B_z}$ is negligible, at most at the order of $1 \times 10^{-6}$ per Tesla (Fig. 4d). The out-of-plane lattice constant change in response to an in-plane magnetic field is also approximately zero (Fig. 4e,f). Similar behavior was demonstrated in $RbV_3Sb_5$ for out-of-plane magnetic field (Fig. 4h). We extracted similarly tiny changes of the *c*-axis lattice constant for other non-magnetic kagome crystals such as $KV_3Sb_5$ and $ScV_6Sn_6$, which are comparable to our result in $CsV_3Sb_5$ and $RbV_3Sb_5$ (Fig.4i, j). It is worth noting that the out-of-plane piezo-magnetic response $\mu_{out-of-plane}^{B_z}$ extracted from the data in Fig. 4c is also absent (inset in Fig. 4c).

**Discussion**

Measuring femtometer-scale crystal distortions generated in solids under external magnetic field using traditional methods can be complicated by the inevitable presence of domains. We establish a procedure to conduct more accurate measurements of the magnetic field driven lattice response at the nanoscale using STM by examining a sequence of rotated scan frames. We note that this is a distinct procedure complementary to the established extraction of spatially varying crystal distortions typically examined at zero magnetic field [60–65] and the temperature-dependent expansion measurements [66]. We applied the method to study magnetic field effects in CDW kagome metals $AV_3Sb_5$. We find that the out-of-plane magnetostriction and piezo-magnetic coefficients are approximately zero. We also reveal that the in-plane lattice deformation is limited by the resolution of current experiments, with the upper bound of about 0.05% for fields of a few Tesla. These results demonstrate that intrinsic magnetic field response of the lattice in $CsV_3Sb_5$ $KV_3Sb_5$ and $RbV_3Sb_5$ in all regions studied here is extremely small in comparison to other experiments on $RbV_3Sb_5$ [50]. Consistent with our previous work, in the regions examined we observe no change in the CDW anisotropy when magnetic field direction is reversed (Fig. 1c). We also note that all experiments are done on sample regions that appear flat, away from obviously buckled regions occasionally seen in layered materials attached in the same geometry [65,67–69]. Future experiments should pursue the investigation of magnetic field coupling to the lattice in such strained regions. It would also be of high interest to apply the experimental method established here to any region where CDW anisotropy appears to be controlled by magnetic field to quantify the magnitude of magnetic field – lattice coupling more accurately.

Importantly, our work cautions the comparison of STM topographs acquired along a single scan direction for the quantitative determination of the atomic lattice constant change, as spurious changes as large as 1% can easily be observed due to experimental artifacts we describe here. The magnitude of these artifacts also appears to fluctuate in different experimental data sets, which we uncovered by examining topographs acquired along different scan directions. Our method mitigates the issue by a meticulous comparison of a series of continuous scans acquired along different crystalline directions, in different magnetic fields, to quantify intrinsic effects. This can be widely applicable to other materials of interest to investigate the coupling of the lattice and externally applied field. We stress that our results do not imply that the CDW phase in $AV_3Sb_5$ does not contain orbital currents or different forms of orbital flux phases [47–49]. However, we demonstrate that the out-of-plane lattice constant is unaffected by magnetic field and that strong coupling of magnetic field and the in-plane lattice structure is not a generic feature of all $AV_3Sb_5$ regions and/or samples studied. Overall our experiments provide further constraints on the origin of time-reversal symmetry breaking in $A$V$_3$Sb$_5$ and establish a procedure to more accurately quantify nanoscale lattice effects in quantum materials.

## Methods

Single crystals of CsV$_3$Sb$_5$ were grown and characterized as described in more detail in Ref. [2] Samples are glued to the sample holder using EPO-TEK H20E silver conducting epoxy and are cured at 175°C for 20 minutes. The cleaving rod is glued to the top of the sample in the same way. We cold-cleave the CsV$_3$Sb$_5$ crystals at around 20 K and quickly insert them into STM for scanning. STM data was acquired using a customized Unisoku USM1300 microscope at approximately 4.5 K. STM tips used were home-made chemically etched tungsten tips, annealed in UHV to bright orange color prior to STM experiments. All STM data shown was acquired at about 4.5 K.

## Acknowledgements

I.Z. gratefully acknowledges the support from the U.S. Department of Energy (DOE) Early Career Award DE-SC0020130 for the development of the method. S.D.W., B.R.O. and A.C.S. gratefully acknowledge support via the UC Santa Barbara NSF Quantum Foundry funded via the Q-AMASE-i program under award DMR-1906325. B.R.O. acknowledges support from the U.S. Department of Energy (DOE), Office of Science, Basic Energy Sciences, Materials Sciences and Engineering Division.

## References

1. Terao, K. & Shimizu, M. Electronic structure and spontaneous volume magnetostriction of antiferromagnetic YMn$_2$. *Phys Lett A* **104**, 113–116 (1984).

2. Ortiz, B. R. *et al.* CsV$_3$Sb$_5$: A Z$_2$ topological kagome metal with a superconducting ground state. *Phys Rev Lett* **125**, 247002 (2020).

3. Ortiz, B. R. *et al.* New kagome prototype materials: discovery of KV$_3$Sb$_5$, RbV$_3$Sb$_5$, and CsV$_3$Sb$_5$. *Phys Rev Mater* **3**, 94407 (2019).

4. Ortiz, B. R. *et al.* Superconductivity in the Z 2 kagome metal KV$_3$Sb$_5$. *Phys Rev Mater* **5**, 34801 (2021).

5. Yang, S.-Y. *et al.* Giant, unconventional anomalous Hall effect in the metallic frustrated magnet candidate, KV$_3$Sb$_5$. *Sci Adv* **6**, eabb6003 (2020).

6. Jiang, Y.-X. *et al.* Unconventional chiral charge order in kagome superconductor KV$_3$Sb$_5$. *Nat Mater* **20**, 1353–1357 (2021).

7. Zhao, H. *et al.* Cascade of correlated electron states in the kagome superconductor CsV$_3$Sb$_5$. *Nature* **599**, 216–221 (2021).

8. Li, H. *et al.* Rotation symmetry breaking in the normal state of a kagome superconductor KV$_3$Sb$_5$. *Nat Phys* **18**, 265–270 (2022).

9. Kenney, E. M., Ortiz, B. R., Wang, C., Wilson, S. D. & Graf, M. J. Absence of local moments in the kagome metal KV$_3$Sb$_5$ as determined by muon spin spectroscopy. *Journal of Physics: Condensed Matter* **33**, 235801 (2021).

10. Chen, H. *et al.* Roton pair density wave in a strong-coupling kagome superconductor. *Nature* **599**, 222–228 (2021).

11. Yu, F. H. *et al.* Unusual competition of superconductivity and charge-density-wave state in a compressed topological kagome metal. *Nat Commun* **12**, 3645 (2021).

12. Yin, Q. *et al.* Superconductivity and Normal-State Properties of Kagome Metal $RbV_3Sb_5$ Single Crystals. *Chinese Physics Letters* **38**, 037403 (2021).

13. Wang, L. *et al.* Anomalous Hall effect and two-dimensional Fermi surfaces in the charge-density-wave state of kagome metal $RbV_3Sb_5$. *Journal of Physics: Materials* **6**, 02LT01 (2023).

14. Yu, F. H. *et al.* Concurrence of anomalous Hall effect and charge density wave in a superconducting topological kagome metal. *Phys Rev B* **104**, L041103- (2021).

15. Li, H. *et al.* Unidirectional coherent quasiparticles in the high-temperature rotational symmetry broken phase of $AV_3Sb_5$ kagome superconductors. *Nat Phys* **19**, 637–643 (2023).

16. Li, H. *et al.* Small Fermi Pockets Intertwined with Charge Stripes and Pair Density Wave Order in a Kagome Superconductor. *Phys. Rev. X* **13**, 31030 (2023).

17. Kang, M. *et al.* Twofold van Hove singularity and origin of charge order in topological kagome superconductor $CsV_3Sb_5$. *Nat Phys* **18**, 301–308 (2022).

18. Hu, Y. *et al.* Rich nature of Van Hove singularities in Kagome superconductor $CsV_3Sb_5$. *Nat Commun* **13**, 2220 (2022).

19. Guo, C. *et al.* Switchable chiral transport in charge-ordered kagome metal $CsV_3Sb_5$. *Nature* **611**, 461–466 (2022).

20. Chen, K. Y. *et al.* Double superconducting dome and triple enhancement of T c in the kagome superconductor $CsV_3Sb_5$ under high pressure. *Phys Rev Lett* **126**, 247001 (2021).

21. Li, H. *et al.* Observation of Unconventional Charge Density Wave without Acoustic Phonon Anomaly in Kagome Superconductors $AV_3Sb_5$ (A= Rb, Cs). *Phys Rev X* **11**, 31050 (2021).

22. Luo, H. *et al.* Electronic nature of charge density wave and electron-phonon coupling in kagome superconductor $KV_3Sb_5$. *Nat Commun* **13**, 273 (2022).

23. Wu, X. *et al.* Nature of Unconventional Pairing in the Kagome Superconductors $AV_3Sb_5$ (A=K,Rb,Cs). *Phys Rev Lett* **127**, 177001 (2021).

24. Xu, H.-S. *et al.* Multiband superconductivity with sign-preserving order parameter in kagome superconductor $CsV_3Sb_5$. *Phys Rev Lett* **127**, 187004 (2021).

25. Zhou, X. *et al.* Origin of charge density wave in the kagome metal $CsV_3Sb_5$ as revealed by optical spectroscopy. *Phys Rev B* **104**, L041101- (2021).

26. Nakayama, K. *et al.* Multiple energy scales and anisotropic energy gap in the charge-density-wave phase of the kagome superconductor CsV$_3$Sb$_5$. *Phys Rev B* **104**, L161112- (2021).

27. Xu, Y. *et al.* Three-state nematicity and magneto-optical Kerr effect in the charge density waves in kagome superconductors. *Nat Phys* **18**, 1470–1475 (2022).

28. Xiang, Y. *et al.* Twofold symmetry of c-axis resistivity in topological kagome superconductor CsV$_3$Sb$_5$ with in-plane rotating magnetic field. *Nat Commun* **12**, 6727 (2021).

29. Wu, Q. *et al.* Simultaneous formation of two-fold rotation symmetry with charge order in the kagome superconductor CsV$_3$Sb$_5$ by optical polarization rotation measurement. *Phys Rev B* **106**, 205109 (2022).

30. Christensen, M. H., Birol, T., Andersen, B. M. & Fernandes, R. M. Theory of the charge density wave in AV$_3$Sb$_5$ kagome metals. *Phys Rev B* **104**, 214513 (2021).

31. Tan, H., Liu, Y., Wang, Z. & Yan, B. Charge density waves and electronic properties of superconducting kagome metals. *Phys Rev Lett* **127**, 46401 (2021).

32. Denner, M. M., Thomale, R. & Neupert, T. Analysis of Charge Order in the Kagome Metal AV$_3$Sb$_5$ (A= K, Rb, Cs). *Phys Rev Lett* **127**, 217601 (2021).

33. Frassineti, J. *et al.* Microscopic nature of the charge-density wave in the kagome superconductor RbV$_3$Sb$_5$. *Phys Rev Res* **5**, L012017 (2023).

34. Li, H., Liu, X., Kim, Y. B. & Kee, H.-Y. Origin of π-shifted three-dimensional charge density waves in the kagomé metal AV$_3$Sb$_5$ (A= Cs, Rb, K). *Phys Rev B* **108**, 75102 (2023).

35. Grandi, F., Consiglio, A., Sentef, M. A., Thomale, R. & Kennes, D. M. Theory of nematic charge orders in kagome metals. *Phys Rev B* **107**, 155131 (2023).

36. Kautzsch, L. *et al.* Incommensurate charge-stripe correlations in the kagome superconductor CsV$_3$Sb$_{5-x}$Sn$_x$. *NPJ Quantum Mater* **8**, 37 (2023).

37. Lin, Y.-P. & Nandkishore, R. M. Complex charge density waves at Van Hove singularity on hexagonal lattices: Haldane-model phase diagram and potential realization in the kagome metals AV$_3$Sb$_5$ (A= K, Rb, Cs). *Phys Rev B* **104**, 45122 (2021).

38. Khasanov, R. *et al.* Time-reversal symmetry broken by charge order in CsV$_3$Sb$_5$. *Phys Rev Res* **4**, 23244 (2022).

39. Farhang, C., Wang, J., Ortiz, B. R., Wilson, S. D. & Xia, J. Unconventional specular optical rotation in the charge ordered state of Kagome metal CsV$_3$Sb$_5$. *Nat Commun* **14**, 5326 (2023).

40. Li, H. *et al.* Discovery of conjoined charge density waves in the kagome superconductor CsV$_3$Sb$_5$. *Nat Commun* **13**, 6348 (2022).

41. Qian, T. *et al.* Revealing the competition between charge density wave and superconductivity in CsV$_3$Sb$_5$ through uniaxial strain. *Phys Rev B* **104**, 144506 (2021).

42. Liang, Z. et al. Three-dimensional charge density wave and surface-dependent vortex-core states in a kagome superconductor $CsV_3Sb_5$. *Phys Rev X* **11**, 31026 (2021).

43. Mielke III, C. et al. Time-reversal symmetry-breaking charge order in a kagome superconductor. *Nature* **602**, 245–250 (2022).

44. Park, T., Ye, M. & Balents, L. Electronic instabilities of kagome metals: saddle points and Landau theory. *Phys Rev B* **104**, 35142 (2021).

45. Nie, L. et al. Charge-density-wave-driven electronic nematicity in a kagome superconductor. *Nature* **604**, 59–64 (2022).

46. Ratcliff, N., Hallett, L., Ortiz, B. R., Wilson, S. D. & Harter, J. W. Coherent phonon spectroscopy and interlayer modulation of charge density wave order in the kagome metal $CsV_3Sb_5$. *Phys Rev Mater* **5**, L111801 (2021).

47. Feng, X., Jiang, K., Wang, Z. & Hu, J. Chiral flux phase in the Kagome superconductor $AV_3Sb_5$. *Sci Bull (Beijing)* **66**, 1384–1388 (2021).

48. Zhou, S. & Wang, Z. Chern Fermi pocket, topological pair density wave, and charge-4e and charge-6e superconductivity in kagomé superconductors. *Nat Commun* **13**, 7288 (2022).

49. Christensen, M. H., Birol, T., Andersen, B. M. & Fernandes, R. M. Loop currents in $AV_3Sb_5$ kagome metals: Multipolar and toroidal magnetic orders. *Phys Rev B* **106**, 144504 (2022).

50. Xing, Y. et al. Optical Manipulation of the Charge Density Wave state in $RbV_3Sb_5$. *arXiv:2308.04128* (2023).

51. Li, H. et al. No observation of chiral flux current in the topological kagome metal $CsV_3Sb_5$. *Phys Rev B* **105**, 45102 (2022).

52. Yothers, M. P., Browder, A. E. & Bumm, L. A. Real-space post-processing correction of thermal drift and piezoelectric actuator nonlinearities in scanning tunneling microscope images. *Review of Scientific Instruments* **88**, 013708 (2017).

53. Jaime, M. et al. Piezomagnetism and magnetoelastic memory in uranium dioxide. *Nat Commun* **8**, 99 (2017).

54. Trainer, C., Abel, C., Bud'ko, S. L., Canfield, P. C. & Wahl, P. Phase diagram of $CeSb_2$ from magnetostriction and magnetization measurements: Evidence for ferrimagnetic and antiferromagnetic states. *Phys Rev B* **104**, 205134 (2021).

55. Ghimire, N. J. et al. Competing magnetic phases and fluctuation-driven scalar spin chirality in the kagome metal $YMn_6Sn_6$. *Sci Adv* **6**, eabe2680 (2024).

56. Wang, Q. et al. Field-induced topological Hall effect and double-fan spin structure with a c-axis component in the metallic kagome antiferromagnetic compound $YMn_6Sn_6$. *Phys Rev B* **103**, 14416 (2021).

57. Zhang, H. et al. Topological magnon bands in a room-temperature kagome magnet. *Phys Rev B* **101**, 100405 (2020).

58. Li, M. *et al.* Dirac cone, flat band and saddle point in kagome magnet $YMn_6Sn_6$. *Nat Commun* **12**, 3129 (2021).

59. Li, H. *et al.* Manipulation of Dirac band curvature and momentum-dependent g factor in a kagome magnet. *Nat Phys* **18**, 644–649 (2022).

60. Zeljkovic, I. *et al.* Strain engineering Dirac surface states in heteroepitaxial topological crystalline insulator thin films. *Nat Nanotechnol* **10**, 849–853 (2015).

61. Watashige, T. *et al.* Evidence for Time-Reversal Symmetry Breaking of the Superconducting State near Twin-Boundary Interfaces in FeSe Revealed by Scanning Tunneling Spectroscopy. *Phys Rev X* **5**, 31022 (2015).

62. Gao, S. *et al.* Atomic-scale strain manipulation of a charge density wave. *Proceedings of the National Academy of Sciences* **115**, 6986–6990 (2018).

63. Ren, Z. *et al.* Nanoscale decoupling of electronic nematicity and structural anisotropy in FeSe thin films. *Nat Commun* **12**, 10 (2021).

64. Sharma, S., Li, H., Ren, Z., Castro, W. A. & Zeljkovic, I. Nanoscale visualization of the thermally driven evolution of antiferromagnetic domains in FeTe thin films. *Phys Rev Mater* **7**, 74401 (2023).

65. Zhao, H. *et al.* Nematic transition and nanoscale suppression of superconductivity in Fe (Te, Se). *Nat Phys* **17**, 903–908 (2021).

66. Crespo, M., Suderow, H., Vieira, S., Bud'ko, S. & Canfield, P. C. Thermal expansion measured by STM in the magnetic superconductor $ErNi_2B_2C$. *Physica B Condens Matter* **378–380**, 471–472 (2006).

67. Soumyanarayanan, A. *et al.* Quantum phase transition from triangular to stripe charge order in $NbSe_2$. *Proceedings of the National Academy of Sciences* **110**, 1623–1627 (2013).

68. Cao, L. *et al.* Two distinct superconducting states controlled by orientations of local wrinkles in LiFeAs. *Nat Commun* **12**, 6312 (2021).

69. Wang, Y. *et al.* Nanoscale strain manipulation of smectic susceptibility in kagome superconductors. *arXiv:2312.06407* (2023).

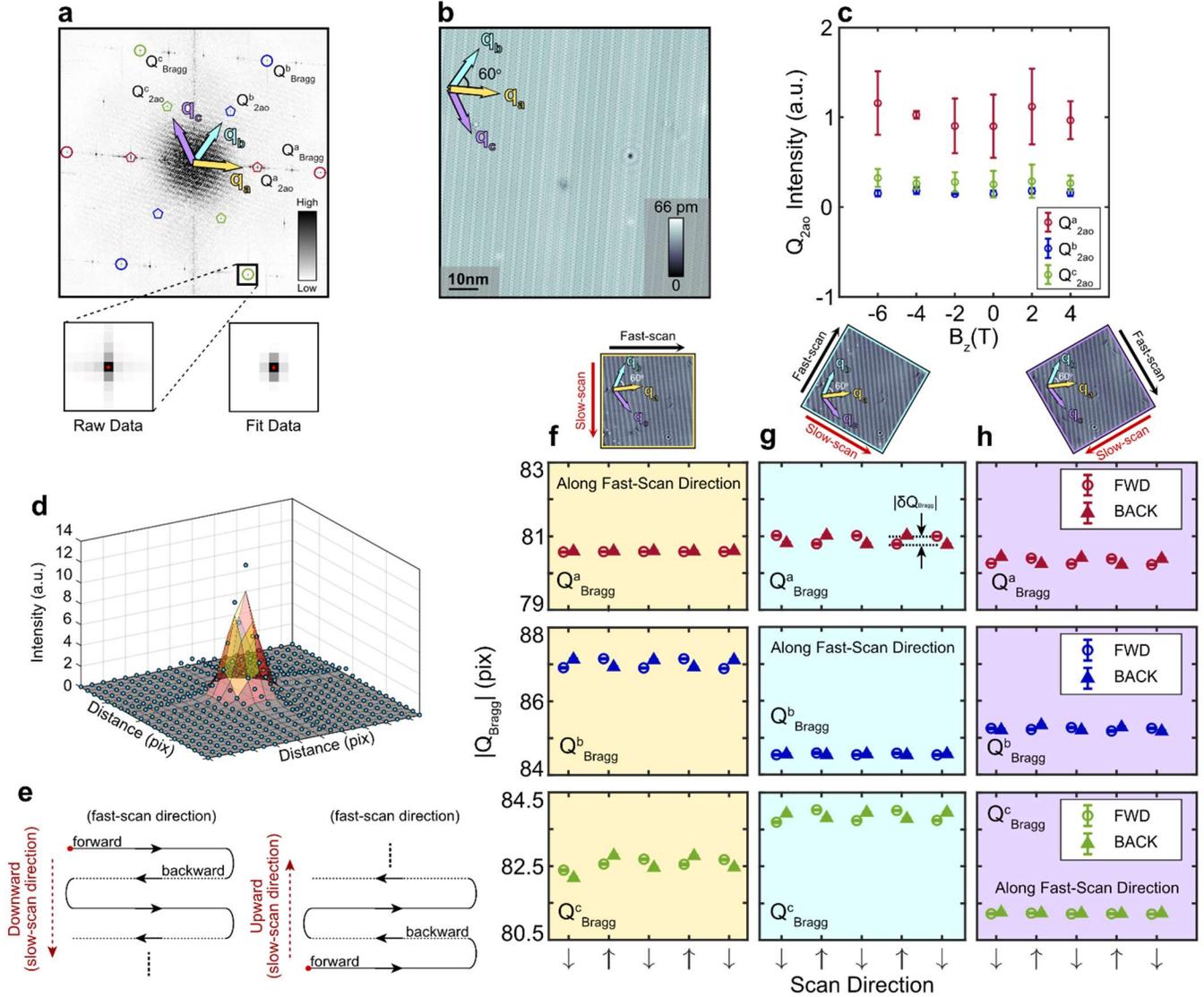

**Figure 1**. **Lattice constant measurements of $CsV_3Sb_5$ from STM topographs as a function of different scan directions.** (a) Fourier transform (FT) of an STM topograph shown in (b) with atomic Bragg peaks (enclosed in circles) and the $2a_o$ charge density wave peaks (enclosed in pentagons) for the three in-plane lattice directions $Q^a_{2a0}$ (yellow), $Q^b_{2a0}$ (blue) and $Q^c_{2a0}$ (purple). Zoomed region shows the raw data around $Q^c_{Bragg}$ (left) and fitted data by a 2D Gaussian (right) with a red dot representing the center of the peak determined from the fitting. The $4a_0$ charge-stripe order is along the $Q^a_{Bragg}$ direction. (b) Representative STM topograph of $CsV_3Sb_5$ Sb termination with the three lattice directions denoted in upper left arrows. Setup conditions were: 100 mV sample bias, 100 pA setpoint current, and fast-scan speed of 30 nm/s. (c) Intensity of the $Q_{2a0}$ charge order peaks along the three directions showing minimal changes with magnetic field applied along the c-axis. Error bars are defined by standard error from comparing 5 measurements of the intensity for each direction and each magnetic field. Data was acquired with fast-scan speed of 10.2 nm/s over a 90 nm x 90 nm (400 pixel x 400 pixel) area. (d) A 3D model demonstrating the Gaussian peak fitting to atomic Bragg peak data. Blue dots represent the raw data; the surface is a mesh of the 2D Gaussian fit. (e) A schematic showing the STM tip scanning

directions to generate topographs. The tip sweeps each line in both forwards and backwards directions (defined as the "fast" scan direction) before moving to the next line in either downward (left) or upward (right) direction, which we term the "slow" scan direction. **(f-h)** Apparent atomic Bragg peak lengths in the units of pixels, extracted from topographs over the same region of the sample but with fast scan directions oriented most closely with each of the three atomic Bragg peaks. Error bars are defined by standard error of the 2D Gaussian fit. Data was acquired with a fast-scan speed of 22.2 nm/s over a 40 nm x 40 nm (256 pixel x 256 pixel) area. The data in **c** is from sample #1, region #1. The data in **f-h** is from sample #1, region #2.

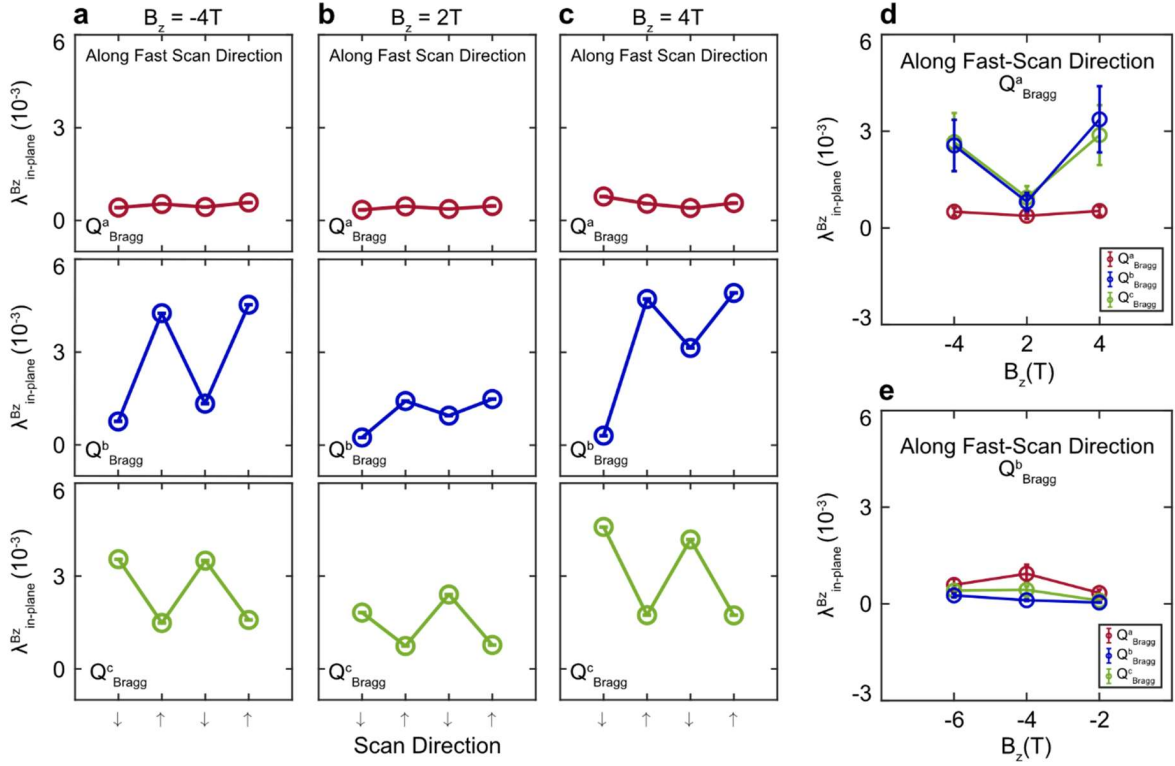

**Figure 2. Spurious detection of atomic Bragg peak changes by magnetic field due to experimental artifacts. (a-c)** Coefficients $\lambda^B_{in-plane} = \frac{|Q^i_{Bragg}(B)| - |Q^i_{Bragg}(0T)|}{|Q^i_{Bragg}(0T)|}$ for a series of upwards and downwards scans at different magnetic fields, with the fast scan direction most closely oriented along the $Q^a_{Bragg}$. Only coefficients along $Q^a_{Bragg}$ show consistent behavior. Error bars are defined by standard error of 2D Gaussian fit. **(d)** Average $\lambda^{B_z}_{in-plane}$ as a function of magnetic field extracted from (a-c), with the fast scan direction along $Q^a_{Bragg}$. As it can be seen, the changes in the atomic Bragg positon $Q^a_{Bragg}$ are minimal and similar across all magnetic fields. **(e)** Average $\lambda^{B_z}_{in-plane}$ as a function of magnetic field over the same area of the sample as in (a-d), but with the fast scan direction along the $Q^b_{Bragg}$ direction. As it can be seen, the change in the atomic Bragg peak positon $Q^b_{Bragg}$ are now minimal and consistent across all fields, in contrast to the, by comparison, enormous values measured in (d), which we conclude to be an artifact. The $4a_0$ charge-stripe order is along the $Q^a_{Bragg}$ direction. All data is from sample #1, region #1 and was acquired using a fast-scan speed of 10.2 nm/s over a 90 nm x 90 nm (400 pix x 400 pix) area. Error bars for (d) and (e) are defined by the following error propagation:

$\delta\lambda = \sqrt{\left(\frac{\delta Q^i_{Bragg}(B)}{|Q^i_{Bragg}(B)|}\right)^2 + \left(\frac{\delta Q^i_{Bragg}(0T)}{|Q^i_{Bragg}(0T)|}\right)^2} \times |\lambda|$ where $\delta Q^i_{Bragg}(B)$ is the standard error of 4 measurements of $Q^i_{Bragg}(B)$ and $|Q^i_{Bragg}(B)|$ is the average value, and $|\lambda|$ is the average magnetostriction observed.

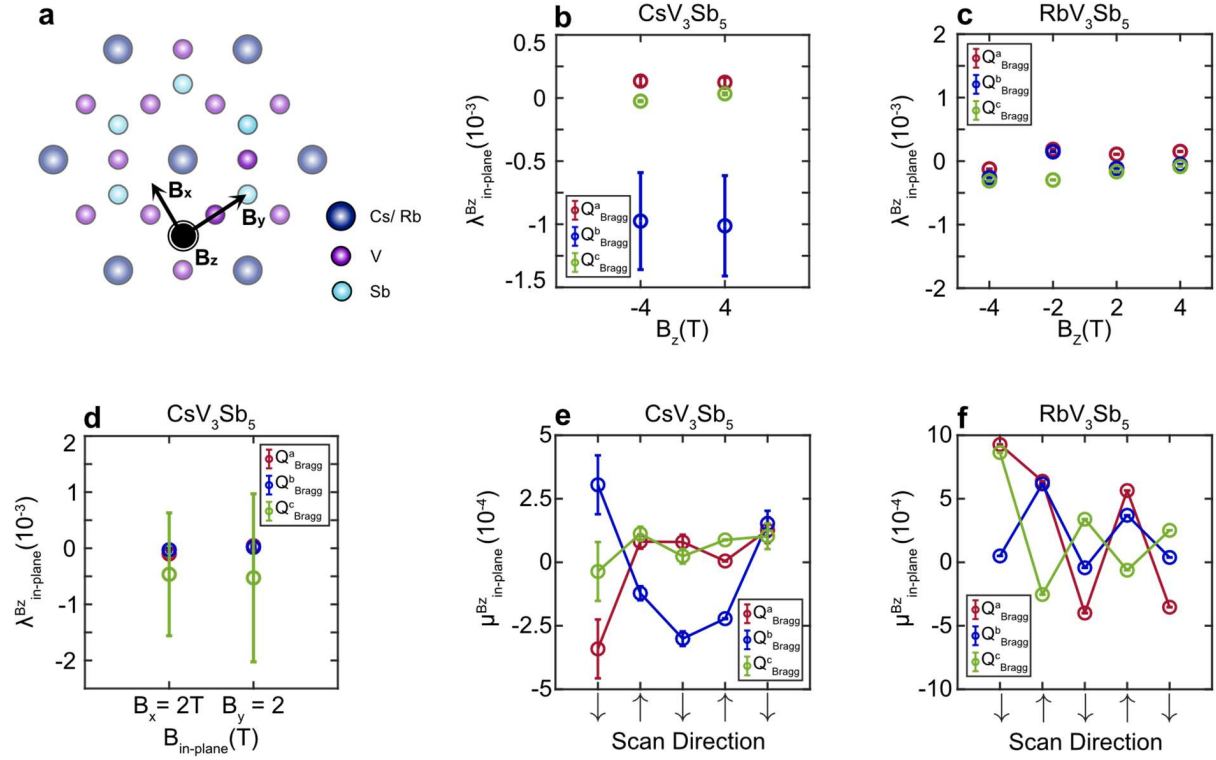

**Figure 3. Quantifying the in-plane lattice response of CsV$_3$Sb$_5$ to magnetic field**. **(a)** A 2D ball model of the crystal structure in the *ab*-plane showing the direction of the 3D vector magnetic field relative to the atomic structure. **(b, c)** In-plane magnetostriction coefficient for (b) CsV$_3$Sb$_5$ and (c) RbV$_3$Sb$_5$ measured by applying an out-of-plane magnetic field $B_z$ ($\lambda^{B_z}_{in-plane}$). **(d)** In-plane magnetostriction coefficient for CsV$_3$Sb$_5$ measured by applying in-plane magnetic field $B_x$ or $B_y$ ($\lambda^{B_{x,y}}_{in-plane}$). Error bars for (b-d) are defined in the same way as those in Fig. 2d,e. **(e, f)** Piezo-magnetic coefficient $\mu^B_{in-plane} = \frac{2(|Q^i_{Bragg}(B)|-|Q^i_{Bragg}(-B)|)}{|Q^i_{Bragg}(B)|+|Q^i_{Bragg}(-B)|}$ for $Q^a_{Bragg}$, $Q^b_{Bragg}$ and $Q^c_{Bragg}$ for (e) CsV$_3$Sb$_5$ and (f) RbV$_3$Sb$_5$ obtained by comparing $B_z$ = 4T and $B_z$ = -4T data for a series of consecutive up and down scans. All data points shown in the figure for a particular lattice direction are extracted from data acquired with a fast-scan direction aligned with that same lattice direction (within a few degrees). The 4$a_0$ charge-stripe order is along the $Q^a_{Bragg}$ direction. Error bars are defined by the following error propagation: $\delta\mu = \sqrt{\left(\frac{\delta Q^i_{Bragg}(-4T)}{|Q^i_{Bragg}(-4T)|}\right)^2 + \left(\frac{\delta Q^i_{Bragg}(4T)}{|Q^i_{Bragg}(4T)|}\right)^2} \times |\mu|$ where $\delta Q^i_{Bragg}(B)$ is the standard error of 2D Gaussian fit and $|Q^i_{Bragg}(B)|$ is the fit value, and $|\mu|$ is the average piezomagnetic response observed. The data in **b** and **e** are from sample #1, region #2 and was acquired using a fast-scan speed of 22.2 nm/s over a 40 nm x 40 nm (256 pixel x 256 pixel) area. The data in **c** is from sample #1, region #1 and was acquired using a fast-scan speed of 30 nm/s over a 30 nm x 30 nm (400 pixel x 400 pixel) area. The data in **d** is from sample #2, region #1 and was acquired using a fast-scan speed of 20 nm/s over a 20 nm x 20 nm (256 pixel x 256 pixel) area.

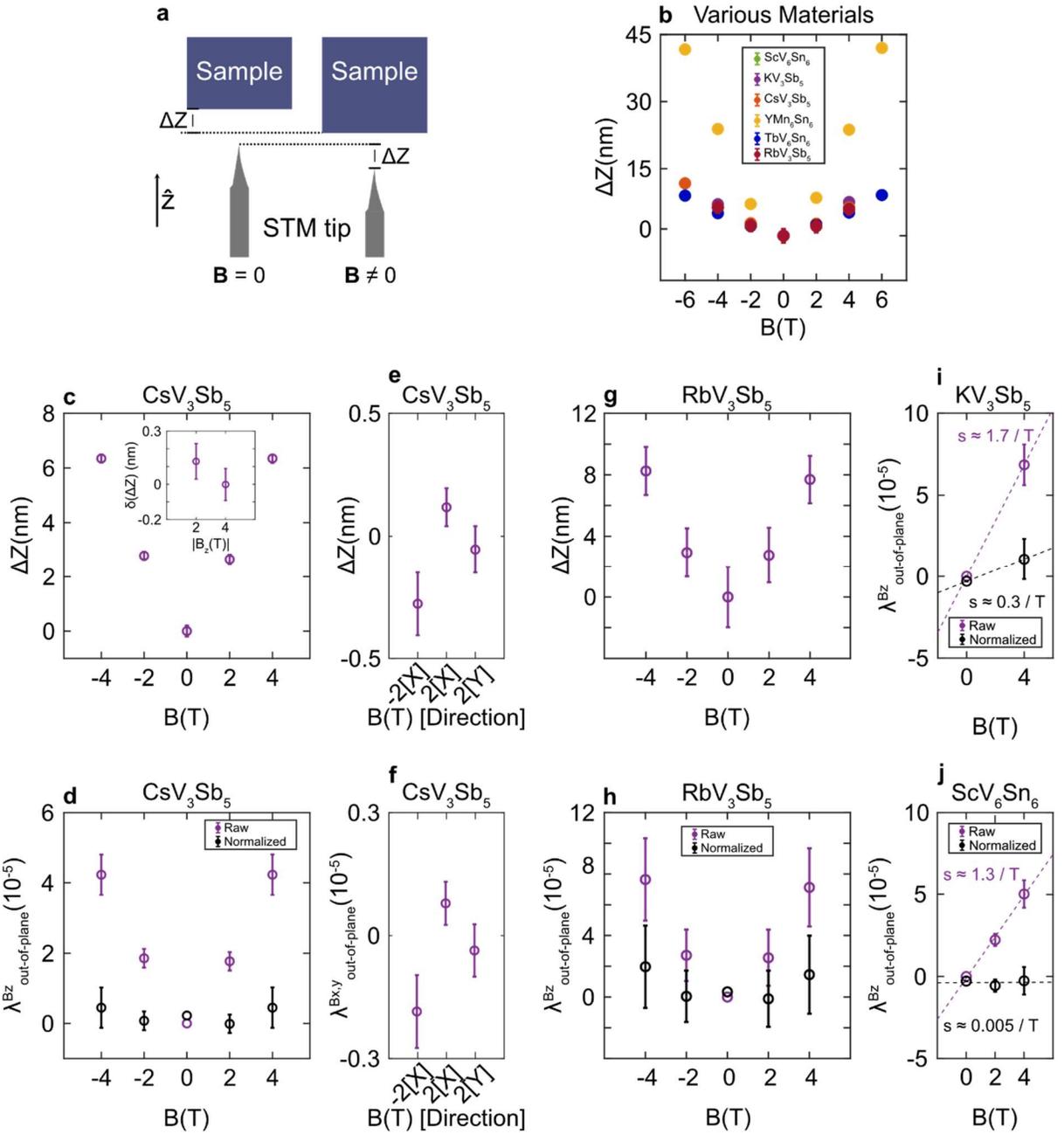

**Figure 4. Out-of-plane lattice constant response to external magnetic field. (a)** Schematic of the measurement. STM tip measures the expansion of the entire sample after magnetic field is applied. The sample in field expands by $\Delta Z$; the feedback loop then adjusts the height of the tip by $\Delta Z$ to achieve the tunneling regime again. **(b)** Summary of $\Delta Z$ vs magnetic field for a range of different materials used. It is worth noting that helical antiferromagnet $YMn_6Sn_6$ shows substantially larger effect compared to all other materials studied. Error bars are defined as the standard error from comparing multiple measurements (number of measurements varies sample to sample). **(c)** Relative change in the height of the STM tip in tunneling ($\Delta Z$) over an identical region of the Sb surface of $CsV_3Sb_5$ as a function of $B_z$ compared to the tip height at 0 T (minus

sign denotes the reversal of the field direction). Error bars are defined as standard error from comparing multiple measurements. **(c inset)** Difference in $\Delta Z$ between negative magnetic field and positive magnetic field of equal magnitudes, demonstrating a piezomagnetic effect less than $10^{-6}$. Error bars are defined by $\sqrt{(\delta Z(-B))^2 + (\delta Z(B))^2}$ where $\delta Z(B)$ is standard error of multiple measurements. **(d)** Coefficients $\lambda_{out-of-plane}^{B_z} = \Delta Z/d$ calculated from (c), taking into account the thickness of the entire sample to be $d = 0.16 \pm 0.02$ mm: assuming the expansion comes from the sample only (purple points), and by accounting for the expansion of the entire sample holder in magnetic field (black symbols). Error bars are defined by the following error propagation:

$$\delta \lambda = \sqrt{\left(\frac{\sqrt{(\delta Z(B))^2 + (\delta Z(0T))^2}}{|Z(B) - Z(0T)|}\right)^2 + \left(\frac{\delta d}{|d|}\right)^2} \times |\lambda| \quad \text{where} \quad \delta Z(B) \text{ is standard error from multiple}$$

measurements, $|Z(B) + Z(0T)|$ is the average from multiple measurements, $\delta d$ is the error in measuring thickness of the material. **(e,f)** Equivalent measurements to those in (c,d), but for an in-plane magnetic fields: $\lambda_{out-of-pla}^{B_{x,y}}$. Error bars are defined equivalently. **(g, h)** Equivalent measurements to those in (c-d) but for the Sb surface of RbV$_3$Sb$_5$. **(i, j)** The c-axis coefficients $\lambda$ for the out-of-plane magnetic field $B_z$ for (g) KV$_3$Sb$_5$ and (h) kagome metal ScV$_6$Sn$_6$ obtained using the same procedure as described in (c,d), which also show a negligible response to magnetic field. Error bars are defined by standard error from multiple measurements. The CsV$_3$Sb$_5$ data in **c, d** is from sample #1, region #1. The CsV$_3$Sb$_5$ data in **e, f** is from sample #2, region #1.